\documentclass[twocolumn,showpacs,preprintnumbers,amsmath,amssymb]{revtex4}
\usepackage{epsfig}
\usepackage{amsmath}
\usepackage{subfigure}
\usepackage{rotating}

\newcommand{\km}{{\langle k \rangle}}

\begin{document}

\title{Chain motifs: The tails and handles of complex networks}

\author{Paulino R. Villas Boas}
\affiliation{Institute of Physics at S\~ao Carlos, University of
S\~ao Paulo, PO Box 369, S\~ao Carlos, S\~ao Paulo, 13560-970
Brazil}

\author{Francisco A. Rodrigues}
\affiliation{Institute of Physics at S\~ao Carlos, University of
S\~ao Paulo, PO Box 369, S\~ao Carlos, S\~ao Paulo, 13560-970
Brazil}

\author{Gonzalo Travieso}
\affiliation{Institute of Physics at S\~ao Carlos, University of
S\~ao Paulo, PO Box 369, S\~ao Carlos, S\~ao Paulo, 13560-970
Brazil}

\author{Luciano da Fontoura Costa}
\affiliation{Institute of Physics at S\~ao Carlos, University of
S\~ao Paulo, PO Box 369, S\~ao Carlos, S\~ao Paulo, 13560-970
Brazil}

\date{\today}

\begin{abstract}

Great part of the interest in complex networks has been motivated by
the presence of structured, frequently non-uniform, connectivity.
Because diverse connectivity patterns tend to result in distinct
network dynamics, and also because they provide the means to identify
and classify several types of complex networks, it becomes important
to obtain meaningful measurements of the local network topology.  In
addition to traditional features such as the node degree, clustering
coefficient and shortest path, motifs have been introduced in the
literature in order to provide complementary description of the
networks connectivity. The current work proposes a new type of motifs,
namely chains of nodes, namely sequences of connected nodes with
degree two. These chains have been subdivided into cords, tails, rings
and handles, depending on the type of their extremities (e.g.\ open or
connected). A theoretical analysis of the density of such motifs in
random and scale free networks is described, and an algorithm for
identifying those motifs in general networks is presented. The
potential of considering chains for network characterization has been
illustrated with respect to five categories of real-world networks
including 16 cases.  Several interesting findings were obtained,
including the fact that several chains were observed in the real-world
networks, especially the WWW, books, and power-grid.  The possibility
of chains resulting from incompletely sampled networks is also
investigated.

\end{abstract}

\maketitle

\section{\label{sec:intr}Introduction}

A large number of interesting dynamic systems can be studied and
modeled by first representing them as networks and then considering
specific dynamic models.  Because the latter depend greatly on the
connectivity of the network, it becomes critical to obtain good
characterizations of the respective connectivity structure.  Such a
characterization is even more important in cases when the dynamics is
not considered, e.g.\ while analyzing a frozen instance of systems
such as the Internet and protein-protein interaction
networks. Therefore, it is hardly surprising that a great deal of
efforts (e.g.\ \cite{Costa:2007:survey}) has been invested in
developing new measurements capable of providing meaningful and
comprehensive characterization of the connectivity structure of
complex networks.

Traditional measurements of the topology of complex networks include
the classical vertex degree and the clustering coefficient (e.g.\
\cite{Newman:2003:survey}).  Both these features are defined for each
vertex in the network and express the connectivity only at the
immediate neighborhood of that reference vertex.  Other measurements
such as the minimum shortest path and betweenness centrality reflect
the connectivity of broader portions of the network.  Hierarchical
measurements (e.g.\ \cite{Costa04:PRL, Costa06:EPJ, Costa06:JSP,
Andrade05:PRL}) such as the hierarchical vertex degree and
hierarchical clustering coefficient, also applicable to individual
reference vertices, have been proposed in order to reflect the
connectivity properties along successive hierarchical neighborhoods
around the reference vertex. Another interesting family of
measurements of the topological properties of complex networks
involves the quantification of the frequency of basic
\emph{motifs} in the network (e.g.\ \cite{ShenOrr:2002, Milo:2002,
Alon:2007:book, Lodato2007}).  Motifs are subgraphs corresponding to
the simplest structural elements found in networks, in the sense of
involving small number of vertices and edges. Examples of motifs
include feed-forward loops, cycles of order three and bi-fans.

The study of chains of nodes in networks has been preliminarily
considered.  Costa~\cite{Costa2004vaf} studied the effect of chains in
affecting the fractal dimension as revealed by dilations along
networks.  Kaiser and Hilgetag~\cite{kaiser2004evn} studied the
vulnerability of networks involving linear chains with an open
extremity.  In another work~\cite{kaiser2004sgr}, they addressed the
presence of this same type of motifs in a sparse model of spatial
network.  More recently, Levnaji\'c and Tadi\'c~\cite{Levnajic}
investigated the dynamics in simple networks including linear chains
of nodes.

Although several measurements are now available in the literature,
their application will always be strongly related to each specific
problem.  In other words, there is no definitive or complete set of
measurements for the characterization of the topology of complex
networks.  For instance, in case one is interested in the community
structures, measurements such as the modularity are more likely to
provide valuable and meaningful information~\cite{Newman04:PRE}.  In
this sense, specific new problems will likely continue to motivate
novel, especially suited, measurements. The reader is referred to the
survey~\cite{Costa:2007:survey} for a more extensive discussion of
measurements choice and applications.

The current work proposes a new, complementary way to characterize the
connectivity of complex networks in terms of a special class of motifs
defined by \emph{chains} of vertices, which are motifs composed by
vertices connected in a sequential way, where the internal vertices
have degree two. These motifs include \emph{cords}, \emph{tails},
\emph{rings} and \emph{handles}. While tails and handles have at least
one extremity connected to the remainder of the network, cords and
rings are disconnected, being composed by groups of vertices connected
in a sequential way. Additional motifs such as two or more handles
connected to the remainder of the network, namely $n$-handles with $n
\ge 2$, can also be defined, but they are not also considered in this
work.

\begin{figure}
  \centerline{\includegraphics[width=0.95\linewidth]{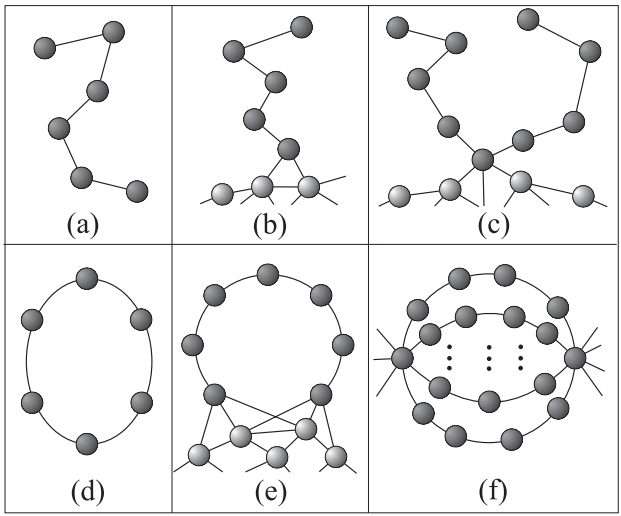}}
  \caption{The chains can be classified into different types, depending
  on the connections among their external vertices. Here is shown six
  types of chains (dark gray vertices): (a) a cord, (b) a tail, (c)
  a two-tail, (d) a ring, (e) a handle and (f) a $n-$handle.}
  \label{Fig:typechains}
\end{figure}

Figure~\ref{Fig:typechains} illustrates six types of chains, namely
(a) a cord, (b) a tail, (c) a two-tail, (d) a ring, (e) a handle and
(f) a $n-$handle. The main difference between the traditional motifs
and those defined and characterized in this article is that the latter
may involve large number of vertices and edges.

The main motivation behind the introduction of the concept of chains
in complex networks provided in this article is that such a structure
is odd in the sense that it can be conceptualized as an edge
containing a series of intermediate vertices which make no
branches. In several aspects, such as in flow, the incorporation of
such intermediate vertices along an edge will imply virtually no
change on the overall dynamics of that substructure of the network. In
other words, the same flow capacity will be offered by either the
isolated edge or its version incorporating a series of intermediate
vertices. Interestingly, vertices with only two neighbors ---
henceforth called \emph{articulations} --- seem to have a rather
distinct nature and role in complex networks, which suggests that they
may have distinct origins.  For instance, as explored further in this
work, articulations seem to appear in networks generated by sequential
processes (e.g.\ word adjacency in books), but can also be a
consequence of incompleteness of the building process of networks.
The latter possibility is experimentally investigated in this work by
considering incompletely sampled versions of network models.

In addition to introducing the concept and a theory of chains and
articulations in complex networks and presenting means for their
identification, the present work also illustrates the potential of the
considering the statistics of cords, tails, and handles for
characterizing real-world networks (social, information,
technological, word adjacency in books, and biological networks).
This article starts by presenting the definition of chains and their
categories (i.e.\ cords, tails, and handles), and proceeds by
developing an analytical investigation of the density of chains in
random and scale free models.  Next, an algorithm for the
identification of such motifs is described, following by a discussion
of the obtained chain statistics.  The application of such a
methodology considers the characterization of real-world complex
networks in terms of chain motifs.

\section{Chains, cords, tails, handles, and rings}
\label{sec:def}

Given a network with $N$ vertices, consider a sequence
$(n_1,n_2,\ldots,n_{m+1})$ of $m+1$ vertices $n_i.$ If the sequence
has the following properties:
\begin{enumerate}
\item There is an edge between vertices $n_i$ and $n_{i+1}$, $1 \le i
\le m$;
\item Vertices $n_1$ and $n_{m+1}$ have degree not equal to 2; and
\item Intermediate vertices $n_i$, $2\le i\le m$, if any, have degree $2$;
\end{enumerate}
we call the sequence a \emph{chain} of length $m$.  Vertices $n_1$ and
$n_{m+1}$ are called the \emph{extremities} of the chain.

Chains can be classified in four categories ($k_{n_i}$ is the degree
of vertex $n_i$):
\begin{description}
\item[Cords] are chains with $k_{n_1}=1$ and $k_{n_{m+1}}=1$.
\item[Handles] are chains with $k_{n_1}>2$ and $k_{n_{m+1}}>2$.
\item[Tails] are chains with $k_{n_1}=1$ and $k_{n_{m+1}}>2$ (or
equivalently $k_{n_1}>2$ and $k_{n_{m+1}}=1$).
\item[Rings] (of length $m$) are sequences $(n_1,n_2,\ldots,n_{m})$ of
  $m$ vertices where the degree of each vertex is $k_{n_i}=2,\,\, 1\le
  n \le m$, $n_i$ is adjacent to $n_{i+1}$ (for $1\le i \le m-1$), and
  $n_m$ is adjacent to $n_1$.
\end{description}

Rings are a special case of chains in which there is no extremities,
and was included in the chain classification only for completeness.

Including the trivial cases with $m=1$, it is easy to see that each
vertex of degree $1$ is at an extremity of a cord or a tail and each
vertex of degree greater than $2$ is at an extremity of a tail or a
handle. Note that the definition of handles includes the degenerate
case where the extremities are the same vertex: $n_1 = n_{m+1}.$

With these definitions and writing $N_C$, $N_H$, $N_T$, and $N_R$ for
the total number of cords, handles, tails, and rings, respectively,
$N(k)$ for the number of vertices of degree $k$ we have:
\begin{eqnarray}
  N(1)  & = & 2 N_C + N_T, \label{eq:deg1}\\
  \sum_{k>2}kN(k) & = & 2 N_H + N_T. \label{eq:degk}
\end{eqnarray}
To evaluate the number of vertices of degree $2$, we introduce the
notation $N_C(m)$ for the number of cords of length $m$, and similarly
$N_H(m)$ for handles, $N_T(m)$ for tails, and $N_R(m)$ for rings. Each
chain of length $m$ has $m-1$ and each ring of length $m$ has $m$
vertices of degree $2$, giving:
\begin{small}
\begin{equation}
  \label{eq:deg2}
  N(2) = \sum_{m=1}^{\infty} \left[ m N_R(m) +
  (m-1)\left( N_C(m)+N_H(m)+N_T(m)\right) \right]
\end{equation}
\end{small}

Isolated vertices (vertices with degree $0$) have no effect on such
structures, and it is considered hereafter that the network has no
isolated nodes.

\begin{figure}
  \centerline{\includegraphics[width=0.6\linewidth]{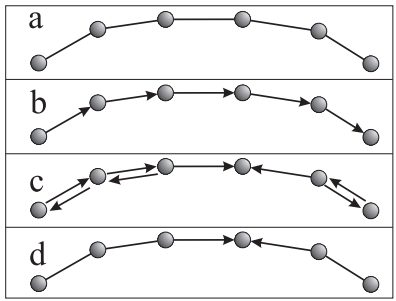}}
  \caption{The chain can be (a) undirected, (b) directed and (c)
  mixed. Mixed chains have arcs in any direction. Note that (c) and
  (d) are equivalent.}
  \label{fig:directions}
\end{figure}

The chains can also be classified according to the nature of its
connections as in Figure~\ref{fig:directions}. In undirected networks,
the chains are said \emph{undirected}
(Figure~\ref{fig:directions}). In directed networks, on the other
hand, the chains can be classified into three types:
\begin{enumerate}
\item \emph{Directed chains} are those whose arcs of inner vertices
follow just one direction, i.e.\ there is a directed path from one
extremity to the other (Figure~\ref{fig:directions}(b)).

\item \emph{Undirected chains} are defined as for undirected networks,
which have undirected arcs between inner vertices
(Figure~\ref{fig:directions}(a)). An undirected arc between vertices
$i$ and $j$ exist if there are an arc from $i$ to $j$ and another from
$j$ to $i$.

\item \emph{Mixed chains} are those with any other combination of arc
directions like in Figure~\ref{fig:directions}(c).
\end{enumerate}

In our analysis we consider just undirect networks, but the extension
for direct networks is straightforward.

\section{Algorithm for chain identification}

\begin{figure*}
  \includegraphics[scale=1]{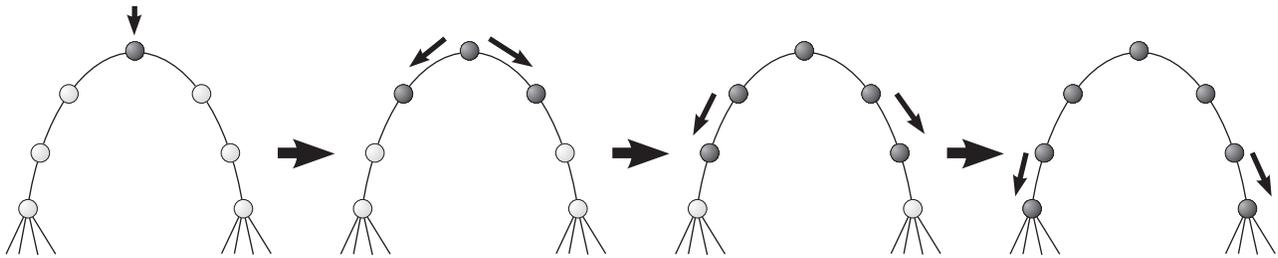}
  \caption{The main steps to identify handles of size greater than 2
  in networks includes: (i) choose a vertex of degree 2 and add it to a
  list (dark gray vertex); (ii) go to its neighbors and also add them
  if they have degree 2; (iii)
  go to the next neighbors, excluding the vertices already added in
  the list, and also add them if they have degree 2; (iv) stop adding
  vertices to the list after finding two vertices of degree greater
  than 2. In this case, the size of the obtained handle is 6. The same
  procedure can also be applied to find cords and tails, but at least
  one extremity should have degree equal to 1.}
  \label{fig:alg}
\end{figure*}

The algorithm to identify chains of vertices includes two steps, one
for finding chains of size greater than 1 and the other for finding
chains of unit size. The first step is illustrated in
Figure~\ref{fig:alg} and described as following:

\begin{small}
\begin{itemize}
  \item input: graph G
  \item output: list containing all chains of size greater than 2
  \item calcule the degree of vertices in G and store them in a list K
  \item Find vertices $i$ such that $k_i \ge 2$, $k_i \in K$, and store them in a
  list Q2
  \item while Q2 is not empty do
  \begin{itemize}
    \item remove a vertex (A) from Q2 and then insert its first
    neighboring vertex (B), A, and its second neighboring vertex (C)
    in a queue P (in this order)
    \item while the first and last elements of P have degree equal to
    2 or are not the same do
    \begin{itemize}
      \item let D be the neighboring node of the first element in P.
            In case D is not already in P, include it into that queue in
            the first position.
      \item if D is in Q2, remove it.
      \item let E be the neighboring node of the last element in P.
            In case E is not already in P, include it into that queue in
            the last position.
      \item if E is in Q2, remove it.
    \end{itemize}
    \item insert P in a list L and clear P
  \end{itemize}
\end{itemize}
\end{small}

The list L contains all chains of size greater than 2.  They can now
be classified into cords, tails, and handles according to the degree
of the first and last element of the corresponding queue.

The second step, required for identifying the chains of unit length,
is as follows:

\begin{small}
\begin{itemize}
  \item input: graph G, list K and list L
  \item output: list of cords, tails, and handles of unit size
  \item find all vertices of degree equal to 1 which were not in L and
  store them in a list Q1
  \item while Q1 is not empty do
  \begin{itemize}
    \item remove a vertex from Q1 and insert it in a queue P
    \item if the neighboring node of A has degree also equal to 1,
    remove it from Q1, insert it in P, and insert P in a list C1
    \item else insert its neighbor in P and insert P in a list T1
  \end{itemize}
  \item include all pairs of connected vertices which are not in L,
  C1 or T1 to a list H1
\end{itemize}
\end{small}

The lists C1, T1, and H1 contain, respectively, all cords, tails, and
handles of unit size in the network.

\section{Statistics} \label{sec:stat}

Consider an ensemble of networks completely determined by the
degree-degree correlations $P(k,k')$~\footnote{For such an ensemble to
be possible, connections from a vertex to itself (self-connections)
and multiple connections between two vertices must be allowed, in
contrast to many network models. Such self- and multiple connections
will be rare provided the network is sufficiently large.}  Given
$P(k,k')$ and the number of vertices in the network, we want to
evaluate the number of each chain type and rings.  The degree
distribution $P(k)$ and the conditional neighbor degree distribution
$P(k'|k)$, i.e.\ the probability that a neighbor of a vertex with
degree $k$ has degree $k'$, are easily computed:
\begin{eqnarray}
  P(k) & = & \frac{\sum_{k'}P(k,k')/k}{\sum_{k',k''}P(k',k'')/k'}, \label{eq:deg}\\
  P(k'|k) & = & \frac{\km P(k,k')}{k P(k)}, \label{eq:cond}
\end{eqnarray}
where $\km = \sum_k k P(k)$ is the average degree of the network.

\subsection{Rings} \label{sec:rings}

For a ring of length $m$, we start at a vertex of degree $2$, go
through $m-1$ vertices of degree $2$ and reach back the original
vertex.  Each transition from a vertex of degree $2$ to the other,
with the exception of the last one that closes the ring, has
probability $P(2|2);$ the closing of the ring requires reaching one of
the vertices of degree $2$ (probability $P(2|2)$) and among them,
exactly the start one (probability $1/(NP(2)$).  If we start from all
vertices of degree $2$, each ring will be counted $m$ times, resulting
in:
\begin{equation}
  \label{eq:rings}
  N_R(m) = \frac{1}{m} P(2|2)^m.
\end{equation}
This expression is valid only for the case of small $m$ and large $N$,
such that the vertices already included in the ring do not affect
significantly the conditional probabilities.  Such an approximation is
used throughout this work.  Note that, under this circumstance, when
computing Eq.~(\ref{eq:deg2}), $N_R(m)$ is of the order of the
approximation error in the expressions of $N_C(m), N_T(m),$ and
$N_H(m).$

\subsection{Cords}
\label{sec:cords}

Starting from a vertex of degree $1$, a cord is traversed by following
through a set of vertices of degree $2$ until reaching a vertex of
degree $1$ that ends the cord.  A cord of length $1$ has no
intermediate vertices; starting in a vertex of degree $1$, the
probability of finding a cord of length 1 is therefore given by
$P(1|1).$ For a cord of length $2$, the edge from the initial vertex
should go through a vertex of degree $2$ before arriving at a new
vertex of degree $1$, giving $P(2|1)P(1|2).$ For lengths greater than
$2$, each new intermediate vertex is reached with probability
$P(2|2)$, and therefore we have $P(2|1)P(2|2)^{m-2}P(1|2)$\footnote{In
these expressions and the following, we assume that the network is
sufficiently large, such that the inclusion of some vertices in the
chain does not affect the probabilities of reaching new vertices in
the next step.} for a cord of length $m$.  Considering that there are
$NP(1)$ vertices of degree $1$ in the network, but only half of them
must be taken as starting vertex to find a cord, we arrive at:
\begin{equation}
  \label{eq:cords}
  N_C(m) = \left\{
        \begin{array}{ll}
          \frac{1}{2}NP(1)P(1|1) & \mbox{if $m = 1$,}\\
          \frac{1}{2}NP(1)P(2|1)P(2|2)^{m-2}P(1|2) & \mbox{if $m > 1$.}\\
        \end{array}
    \right.
\end{equation}

\subsection{Tails}
\label{sec:tails}

The number of tails can be computed similarly. We need either to start
at a vertex with degree $1$ and reach a vertex of degree greater than
$2$ or vice versa; only one of these possibilities must be
considered. We arrive at:
\begin{equation}
  \label{eq:tails}
  N_T(m) = \left\{
        \begin{array}{ll}
          NP(1)P(>2|1) & \mbox{if $m = 1$,}\\
          NP(1)P(2|1)P(2|2)^{m-2}P(>2|2) & \mbox{if $m > 1$,}\\
        \end{array}
    \right.
\end{equation}
where the notation $P(>2|k) = \sum_{k'>2}P(k'|k)$ is used.

\subsection{Handles}
\label{sec:handles}

A handle starts in a vertex of degree $k>2$ and ends in a vertex of
degree $k'>2.$ Starting from one of the $NP(k)$ vertices of degree
$k>2$ of the network, there are $k$ possibilities to follow a chain,
each characterized by a sequence of vertices of degree $2$ until
reaching a vertex of degree $k'>2.$ This gives a total of
$NkP(k)P(>2|k)$ handles of length $1$ and
$NkP(k)P(2|k)P(2|2)^{m-2}P(>2|2)$ handles of length $m>1.$ Summing up
for all values of $k>2$, using $\sum_{k}kP(k)P(k'|k) = k'P(k'),$ which
can be deduced from relations~(\ref{eq:deg}) and~(\ref{eq:cond}), and
considering that each handles is counted twice when starting from all
nodes of degree greater than 2, we have:
\begin{widetext}
\begin{equation}
  \label{eq:handles}
  N_H(m)  = \left\{
        \begin{array}{ll}
          \frac{1}{2}N\left\{\km -  
            P(1)[2-P(1|1)-P(2|1)]- P(2)[4-P(1|2)-P(2|2)]\right\}
          & \mbox{if $m = 1$,} \\
          \frac{1}{2}N[2P(2)-P(1)P(2|1)-2P(2)P(2|2)]P(2|2)^{m-2}P(>2|2)
          & \mbox{if $m > 1$.}\\
        \end{array}
    \right.
\end{equation}
\end{widetext}
Using Equations~(\ref{eq:cords}), (\ref{eq:tails}),
and~(\ref{eq:handles}) we have
\begin{displaymath}
  \sum_{m=1}^{\infty} \left[ (m-1) \left( N_C(m)+N_H(m)+N_T(m)\right)
  \right] = N(2).
\end{displaymath}
Comparing this result with Equation~(\ref{eq:deg2}) we see that the
rings are already counted in the number of chains, as hinted in the
end of Section~\ref{sec:rings}.  This happens because, while computing
the probability of chains, we ignore the fact that the presence of
rings decreases the number of possible chains.  For a large enough
network, the number of rings should be small compared with the number
of the other structures, validating the approximation.

Note that all expressions are proportional to $P(2|2)^m$, and
therefore large chains should be exponentially rare, if they are not
favored by the network growth.

\section{Theoretical analysis for uncorrelated networks} \label{sec:uncorr}

For uncorrelated networks, where the degree at one side of an edge is
independent of the degree at the other side of the edge, $P(k,k')$ can
be factored as
\begin{equation}
  \label{eq:pkkuncorr}
  P(k,k') = \frac{kP(k)k'P(k')}{\km^2}.
\end{equation}
The conditional probability is simplified to
\begin{equation}
  \label{eq:conduncorr}
  P(k'|k) = \frac{k'P(k')}{\km}.
\end{equation}
Using this last expression, we have for uncorrelated networks
\begin{eqnarray}
  \label{eq:nruncorr}
  N_R(m) & = & \frac{1}{m} \left[\frac{2P(2)}{\km}\right]^m\\
  \label{eq:ncuncorr}
  N_C(m) & = & \frac{2^{m-2}NP(1)^2 P(2)^{m-1}}{\km^m}  \\
  \label{eq:ntuncorr}
  N_T(m) & = & NP(1)\left[\frac{2P(2)}{\km}\right]^{m-1}
  \alpha\\
  \label{eq:nhuncorr}
  N_H(m) & = & \frac{N\km}{2}\left[ \frac{2P(2)}{\km}\right]^{m-1}
  \alpha^2.
\end{eqnarray}
where $\alpha = \left[1 - \frac{P(1)}{\km} - \frac{2P(2)}{\km}\right]$.

\subsubsection{Erd\H{o}s-R\'{e}nyi networks} \label{sec:er}

\begin{figure}
  \includegraphics[scale=0.75]{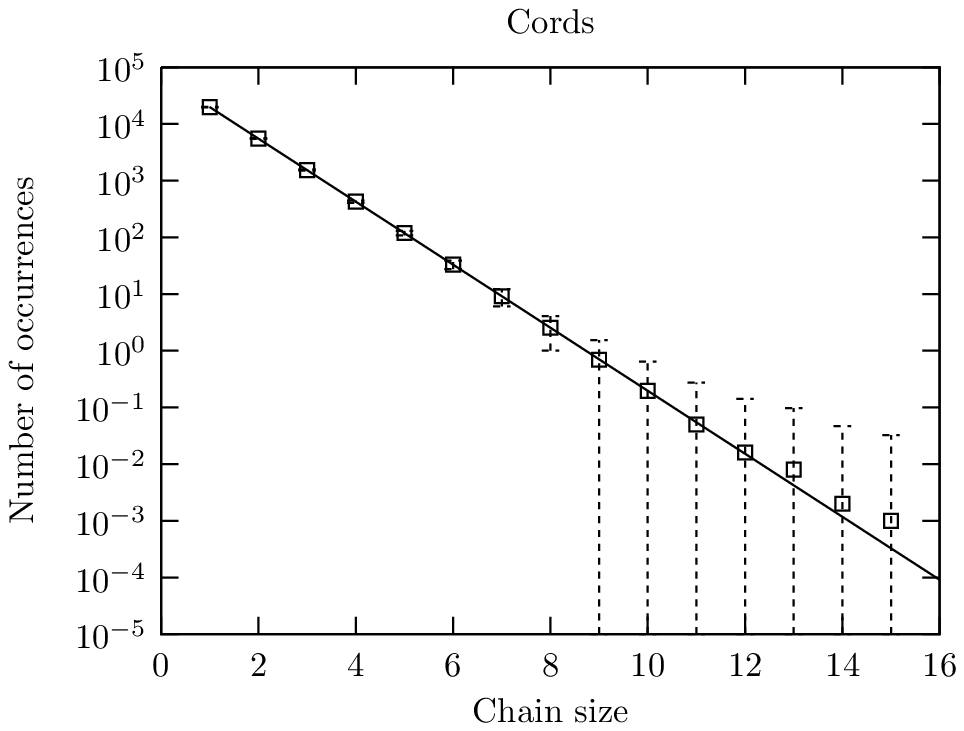} (a) \vspace{0.2cm}\\
  \includegraphics[scale=0.75]{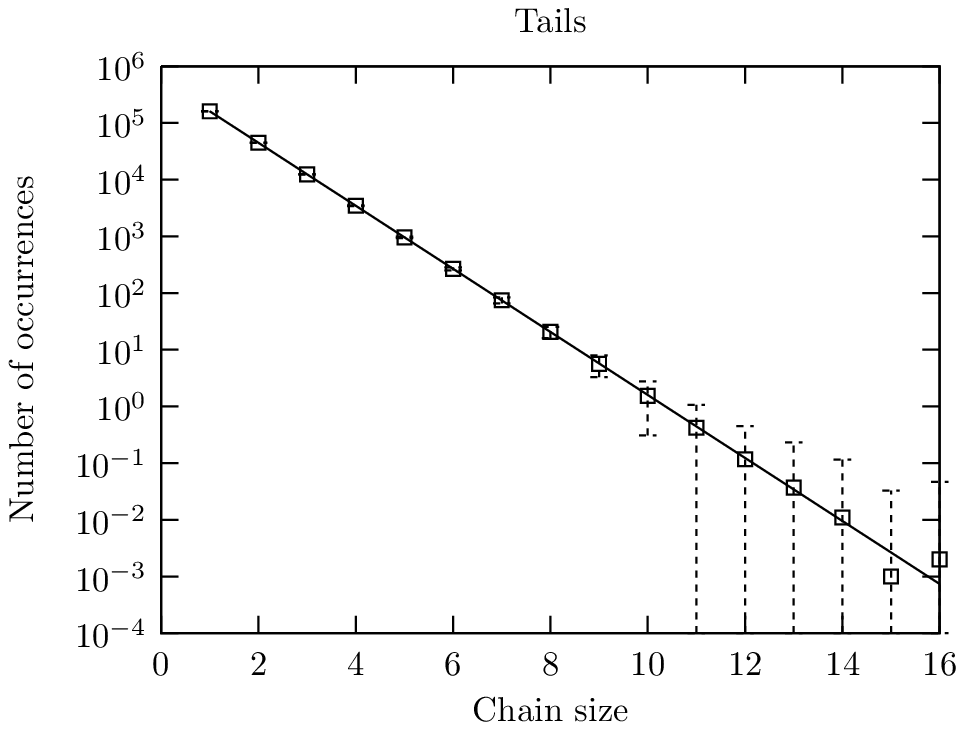} (b) \vspace{0.2cm}\\
  \includegraphics[scale=0.75]{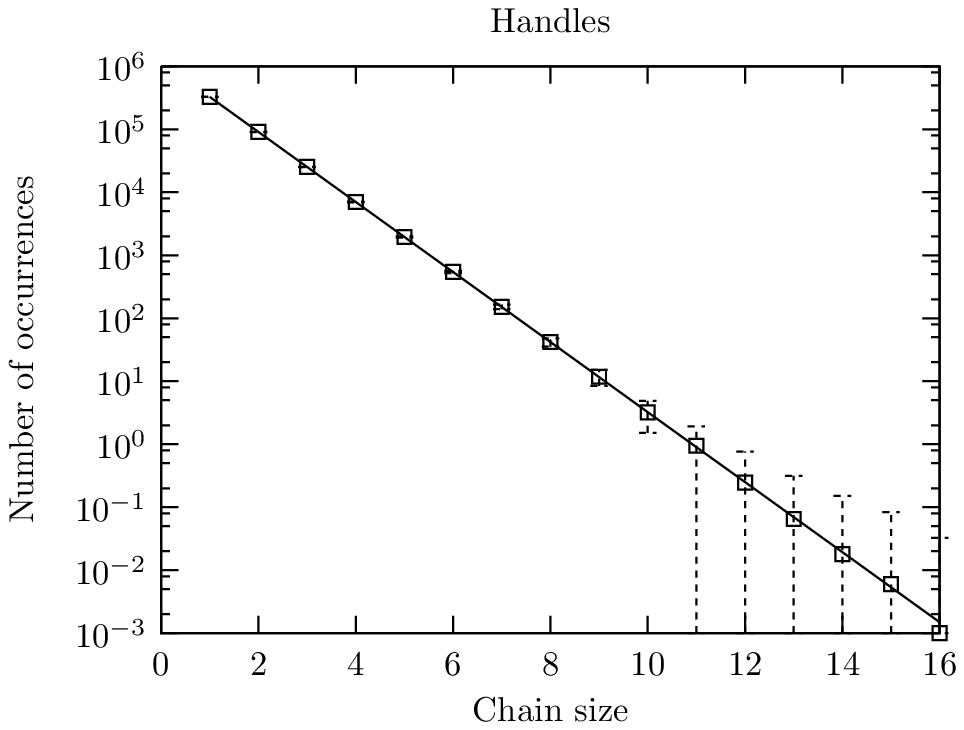} (c)\\
  \caption{Number of cords (a), tails (b), and handles (c) of
    different sizes in the model with Poisson degree distribution.
    The points are the averaged measured values (each of the error bars
    corresponds to one standard deviation), the lines are the values computed
    analytically. Note that the abrupt increase of the width of the
    error bars is a consequence of the logarithmic scale.}
  \label{fig:poisson}
\end{figure}

Erd\H{o}s-R\'{e}nyi networks have no degree correlations and a
Poissonian degree distribution:
\begin{equation}
  \label{eq:pker}
  P(k) = \frac{e^{-\km}\km^k}{k!}.
\end{equation}
This gives the following expressions for the number of rings, cords,
tails and handles:
\begin{eqnarray}
  \label{eq:ernr}
  N_R(m) & = & \frac{\km^m e^{-m \km}}{m}\\
  \label{eq:ernc}
  N_C(m) & = & \frac{N}{2} \km^m e^{-(m+1)\km}\\
  \label{eq:ernt}
  N_T(m) & = & N \km^m e^{-(m+1)\km} \varepsilon\\
  \label{eq:ernh}
  N_H(m) & = & \frac{N}{2} \km^m e^{-(m+1)\km} \varepsilon^2
\end{eqnarray}
where $\varepsilon=\left(e^\km - \km - 1\right)$.
Figure~\ref{fig:poisson} shows the comparison of the results for
networks with $N=10^6$ vertices and $L=972\,941$ edges (this number of
edges was chosen to give the same average degree as for the scale-free
network discussed below).  A total of 1\,000 realizations of the
model were used to compute the averages and standard deviations.

\subsubsection{Scale-free networks} \label{sec:sf}

We now proceed to uncorrelated scale-free networks with degree
distribution given as
\begin{equation}
  \label{eq:pksf}
  P(k) = \frac{k^{-\gamma}}{\zeta(\gamma)},
\end{equation}
where $\gamma$ is the power law coefficient and $\zeta(x)$ is the
Riemann zeta function.  This distribution describes a strictly
scale-free network, with the power law valid for all values of $k$ and
a minimum $k_{\mathrm{min}} = 1.$ The results are therefore not
directly applicable to scale-free real networks or models.  The
average degree is $\km = \zeta(\gamma-1)/\zeta(\gamma).$ The resulting
expressions are:
\begin{eqnarray}
  \label{eq:sfnr}
  N_R(m) & = & \frac{2^{-m(\gamma-1)}}{m\zeta(\gamma-1)^m}\\
  \label{eq:sfnc}
  N_C(m) & = & \frac{N}{2} \frac{2^{-(m-1)(\gamma-1)}}
                                {\zeta(\gamma)\zeta(\gamma-1)^m}\\
  \label{eq:sfnt}
  N_T(m) & = & N
  \frac{2^{-(m-1)(\gamma-1)}}{\zeta(\gamma)\zeta(\gamma-1)^m}
  \beta \\
  \label{eq:sfnh}
  N_H(m) & = & \frac{N}{2}
  \frac{2^{-(m-1)(\gamma-1)}}{\zeta(\gamma)\zeta(\gamma-1)^m}
  \beta^2
\end{eqnarray}
where $\beta=\left[\zeta(\gamma-1)-1-2^{-(\gamma-1)}\right]^2$.

\begin{figure}
  \includegraphics[scale=0.75]{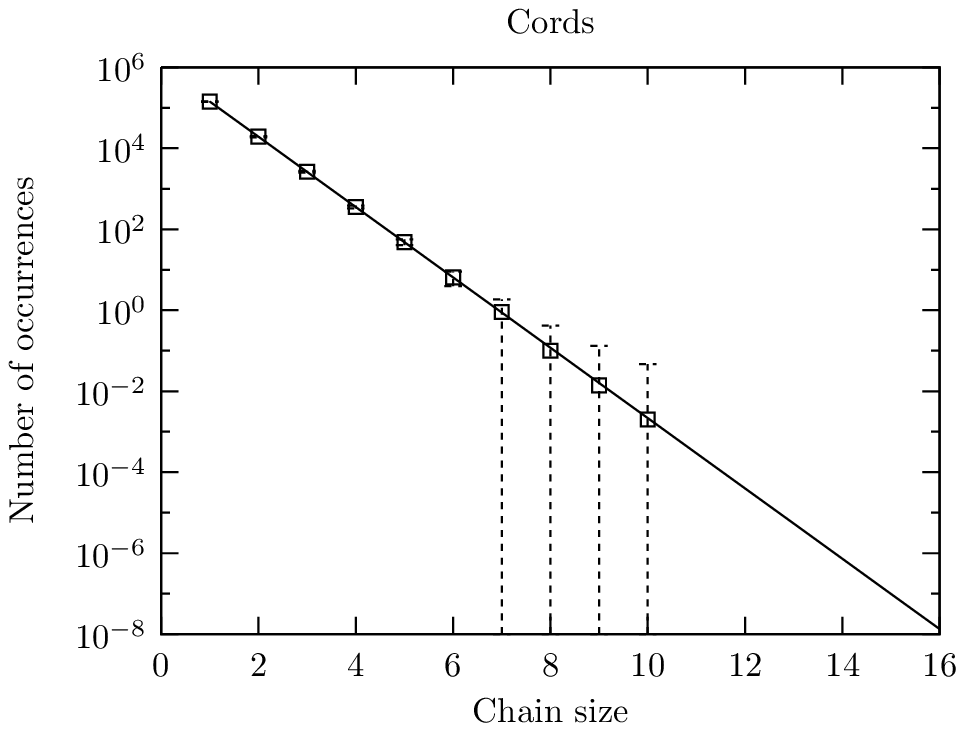} (a) \vspace{0.2cm}\\
  \includegraphics[scale=0.75]{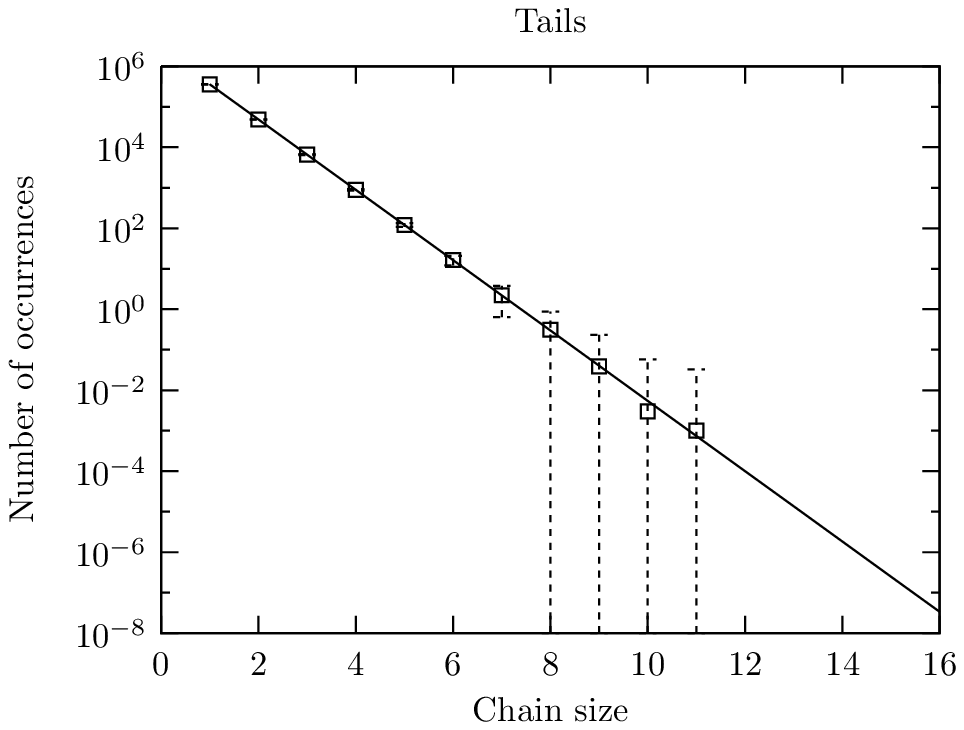} (b) \vspace{0.2cm}\\
  \includegraphics[scale=0.75]{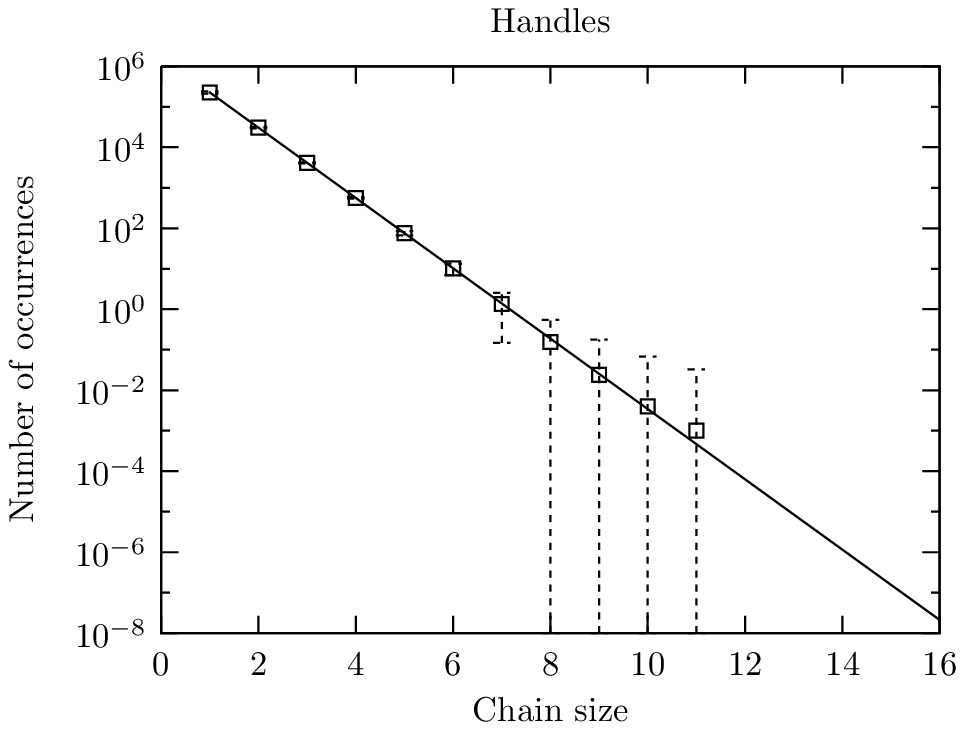} (c)\\
  \caption{Number of cords (a), tails (b), and handles (c) of
    different sizes in the model with scale-free degree distribution.
    The points are the averaged measured values (each of the
    error bars corresponds to one standard deviation), the lines
    are the values computed analytically.}
  \label{fig:sf}
\end{figure}

Figure~\ref{fig:sf} shows the comparison of the results for networks
with $N=10^6$ vertices and $\gamma=2.5$.  A total of 1\,000
realizations of the model were used to compute the averages and
standard deviations.  A comparison with Figure~\ref{fig:poisson} shows
that the Poisson degree distribution with the same average degree
presents larger chains.  This is due to the relation between the
constants in the exponential dependency with $m$: $\langle k
\rangle/e^{\langle k \rangle} \approx 0.278$ for the Poisson model and
$2^{1-\gamma}/\zeta(\gamma-1)\approx 0.135$ for the scale-free model.

The results presented in this section addressed the issue of
validating the theory for analytical models. In
Section~\ref{sec:uncorr}, we will evaluate the theory while
considering real-world networks.

\section{\label{sec:netdata}Real-world networks}

It is known that networks belonging to the same class may share
similar structural properties~\cite{Milo:2002,Newman:2003}. So, to
study the presence of handles in networks, we considered five types of
complex networks, namely social networks, information networks, word
adjacency networks in books, technological networks, and biological
networks.

\subsection{Social networks}

Social networks are formed by people or group of people (firms, teams,
economical classes) connected by some type of interaction, as
friendship, business relationship between companies, collaboration in
science and participation in movies or sport
teams~\cite{Newman:2003:survey}, to cite just a few examples. Below we
describe the social networks considered in our analysis.

\begin{trivlist}

\item \textbf{Scientific collaboration networks} are formed by
scientists who are connected if they had authored a paper together. In
our investigations, we considered the astrophysics collaboration
network, the condensed matter collaboration network, the high-energy
theory collaboration network, all collected by Mark Newman from
\texttt{http://www.arxiv.org}, and the scientific collaboration of
complex networks researchers, also compiled by Mark Newman from the
bibliographies of two review articles on networks (by
Newman~\cite{Newman:2003:survey} and Boccaletti et
al.~\cite{Boccaletti06}).  The astrophysics collaboration network is
formed by scientists who post preprints on the astrophysics archive,
between the years 1995 and 1999~\cite{Newman-PNAS01}. The condensed
matter collaboration network, on the other hand, is composed by
scientist posting preprints on the condensed matter archive from 1995
until 2005~\cite{Newman-PNAS01}.  Finally, the high-energy theory
collaboration network is composed by scientists who posted preprints
on the high-energy theory archive from 1995 until
1999~\cite{Newman00:PRE64:I,Newman00:PRE64:II}.

\end{trivlist}

\subsection{Information networks}

\begin{trivlist}

\item \textbf{Roget's Thesaurus network} is constructed associating
each vertex of the network to the one of the 1022 categories in the
1879 edition of Peter Mark Roget's Thesaurus of English Words and
Phrases, edited by John Lewis Roget~\cite{Roget82}. Two categories $i$
and $j$ are linked if Roget gave a reference to $j$ among the words
and phrases of $i$, or if such two categories are directly related to
each other by their positions in Roget's book~\cite{Roget82}.  Such
network is available at Pajek datasets~\cite{pajek-data}.

\item \textbf{Wordnet} is a semantic network which is often used as a
form of knowledge representation. It is a directed graph consisting of
concepts connected by semantic relations. We collected the network
from the Pajek datasets~\cite{pajek-data}.

\item \textbf{The World Wide Web} is a network of Web pages belonging
to nd.edu domain linked together by hyperlinks from one page to
another~\cite{Albert99:Nature}. The data considered in our paper is
available at the Center for Complex Network Research~\cite{CCNR}
\end{trivlist}

\subsection{Word adjacency in books}

Word adjacency in books can be represented as a network of words
connected by proximity~\cite{Antiqueira2007}. A directed edge is
established between two words that are adjacent and its weight is the
number of times the adjacent words appear in the text. Before
constructing a network, the text must be preprocessed. All stop words
(e.g.\ articles, prepositions, conjunctions, etc) are removed, and the
remaining words are lemmatized~\cite{Antiqueira2007}. In our analysis,
we considered the books: David Copperfield by Charles Dickens, Night
and Day by Virginia Woolf, and On the Origin of Species by Charles
Darwin compiled by Antiqueira~\emph{et al.}~\cite{Antiqueira2006}.

\subsection{Technological networks}

\begin{trivlist}
\item \textbf{Internet} or the autonomous systems (AS) network is a
collection of IP networks and routers under the control of one entity
that presents a common routing policy to the Internet.  Each AS is a
large domain of IP addresses that usually belongs to one organization
such as a university, a business enterpriser, or an Internet Service
Provider. In this type of networks, two vertices are connected
according to BGP tables. The considered network in our analysis was
collected by Newman in July, 2006~\cite{Newman:data}.

\item \textbf{The US Airlines Transportation Network} is formed by US
airports in 1997 connected by flights. Such network is available at
Pajek datasets~\cite{pajek-data}.

\item \textbf{The Western States Power Grid} represents the topology
of the electrical distribution grid~\cite{Watts:1998}. Vertices
represent generators, transformers and substations, and edges the
high-voltage transmission lines that connect them.

\end{trivlist}

\subsection{Biological networks}

Some biological systems can be modeled in terms of networks as the
brain, the genetic interaction and the interaction between proteins.

\begin{trivlist}

\item \textbf{The neural network of \emph{Caenorhabditis elegans}} is
composed by neurons connected according to
synapses~\cite{White86,Watts:1998}.

\item \textbf{Transcriptional Regulation Network of the Escherichia
coli} is formed by operons (an operon is a group of contiguous genes
that are transcribed into a single mRNA molecule). Each edge is
directed from an operon that encodes a transcription factor to another
operon which is regulated by that transcription factor. This kind of
network plays an important role in controlling gene
expression~\cite{ShenOrr:2002}.

\item \textbf{The protein-protein interaction network of
\emph{Saccharomyces cerevisiae}} is formed by proteins connected
according to identified directed physical interactions~\cite{Jeong01}.

\end{trivlist}

\section{\label{sec:appl}Results and Discussion}

We analyzed the real-world networks by comparing their number of
cords, tails, and handles with random networks generated by the
rewiring procedure as described in~\cite{milo2003} and with the
theory proposed in Section~\ref{sec:stat}.

\subsection{Comparison between real-world networks and their
randomized counterparts}

For each considered real-world network, we generated 1\,000 randomized
versions (100 for WWW) by the rewiring process described
in~\cite{milo2003}. The generated networks have the same degree
distribution as the original, but without any degree-degree
correlation. In order to compare the chain statistics obtained for the
real-world and the respective randomized versions, we evaluated the
Z-score values for each size of the cords, tails, and handles. The
Z-score is given by,

\begin{equation}
Z = \frac{X_{\mathrm{Real}}-\langle X \rangle}{\sigma},
\end{equation}
where $X_{\mathrm{Real}}$ is the number of cords, tails, or handles
with a specific size of the original (real-world) analyzed network,
and $\langle X \rangle$ and $\sigma$ are, respectively, the average
and the standard deviation of the corresponding values of its
randomized counterparts. A null value of the Z-score indicates that
there is no statistical difference between the number of occurrences
of cords, tails, or handles in the considered network and in its
randomized versions.

\begin{figure*}
  \includegraphics[width=0.99\linewidth]{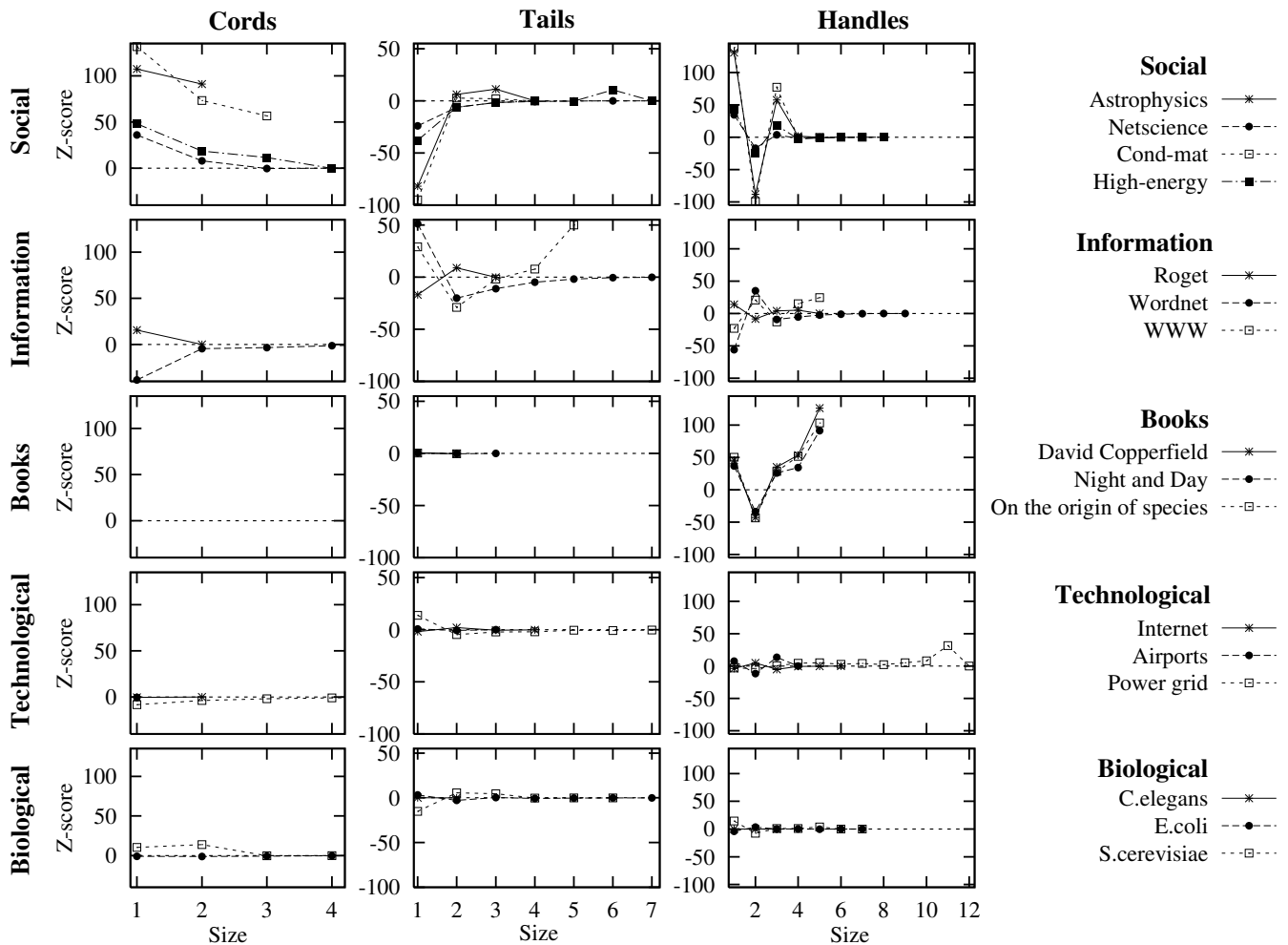}
  \caption{Z-scores of the number of cords, tails, and handles for each
  size.    The number of generated random networks was 1\,000 for all
  considered networks, except for WWW, which was 100 (because of the
  substantially larger size of this network).}
  \label{fig:zscore}
\end{figure*}

The results of the Z-scores for all considered networks can be seen in
Figure~\ref{fig:zscore}. The cases in which the Z-score values are not
defined ($\sigma = 0$) were disconsidered.

The majority of results presented in Figure~\ref{fig:zscore} can be
explained by the fact that the rewiring process tends to make uniform
the distribution of cords size, tails and handles. In this way, the
excess of these structures on the real networks will reduce in the
random counterparts. For instance, if a network have many large
handles, its random version will present few large handles but many
small ones. The next discussion will not take into account the shape
of the distribution of chains, but just the most important results.

In the case of collaboration networks, there is a large quantity of
cords. This fact suggests that researchers published papers with just
one, two or three other scientists. Cords may appear because many
researchers can publish in other areas and, therefore, such papers are
not included in the network. If other research areas had been
considered, this effect could not occur and the number of small cords
would be less significant. Thus, the presence of cords in
collaboration networks can be the result of database incompleteness.
Another possible cause of cords in such networks concerns the
situations of authors which publish only among themselves.

The information networks do not present a well defined patterns as
observed in collaboration network. The Roget thesaurus network is
different from the others, but the results obtained for such a network
are not expressive enough to be discussed. Important to note that in
the Wordnet and WWW, there is a large occurrence of tails of size
one. In the case of Wordnet, this happen because specific words has
connections with more common words which has connections with the
remainder of the network. In the case of WWW, this structure is a
consequence of characteristic url documents which have just one
link. In addition to small tails, the WWW have long tails and handles.
This fact can be associated to the way in which the network were
constructed, by considering a \emph{web
crawler}~\cite{Albert99:Nature} --- a program designed to visit url
documents inside a given domain and get links between them in a
recursive fashion. When pages are visited by the crawler, the wandered
path can originate chains. If the program is not executed by a long
time interval, long chains can appear. Thus, this effect can be
resulting of incomplete sampling (see
Subsection~\ref{sec:incompleteness}). Besides, as the process of
network construction is recursive, isolated components does not occurs
in the database and therefore there are no cords and rings.

The books adjacency networks presents a characteristic pattern of
chains: no cords, the same quantity of tails of sizes 1, 2 and 3 as
observed in the random counterparts, and many handles of size 1, 3, 4
and 5.  The increasing in the quantity of handles of size 2 in random
versions are consequence of the fact that when the rewiring process
are performed, many handles of size one can be put together. This
fact explain why book networks present more handles of size one than
in random counterparts.  On the other hand, the long handles are
consequence of the sequential process considered to obtain the
network.

In technological networks, the chain patterns are more significant in
power grid. This networks present a high quantity of tails of size one
and handles of size 11. While the first occurrence appear to be
related to the geographical effect, where new vertices needed to cover
a new region tend to connect with the near vertices, the second can be
resulting of geographical constraints (e.g.\ the transmissors may be
allocated in a strategic way in order to contour a mountain, lake or
other geographical accidents).

The results obtained for biological networks are not so
expressive. However, the protein interaction network of the yeast
\emph{S. cerevisiae} have many cords of size one and two. The presence
of small cords in this networks is a consequence of isolated chains of
proteins which interact only with a small number of other
proteins. This fact can be due to incompleteness~\cite{han2005est},
where many real connections may not be considered, or high specialized
proteins, which lost many connections because the mutation process ---
protein interaction networks evolve from two basic process:
duplication and mutation~\cite{Vazquez03:complexus}.

\subsection{Theoretical analysis of the real-world networks}

\begin{figure*}
  \subfigure[\,Number of
  cords.]{\includegraphics[width=0.54\linewidth]{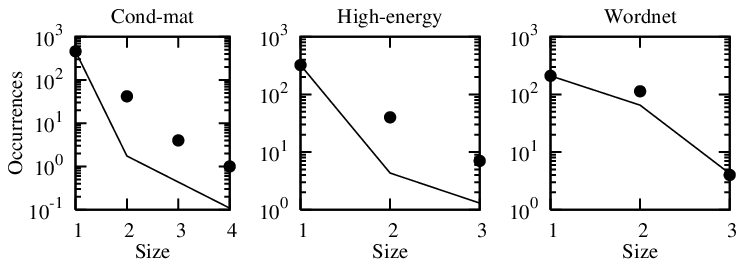}}
  \vspace{0.3cm} \subfigure[\,Number of
  tails.]{\includegraphics[width=0.9\linewidth]{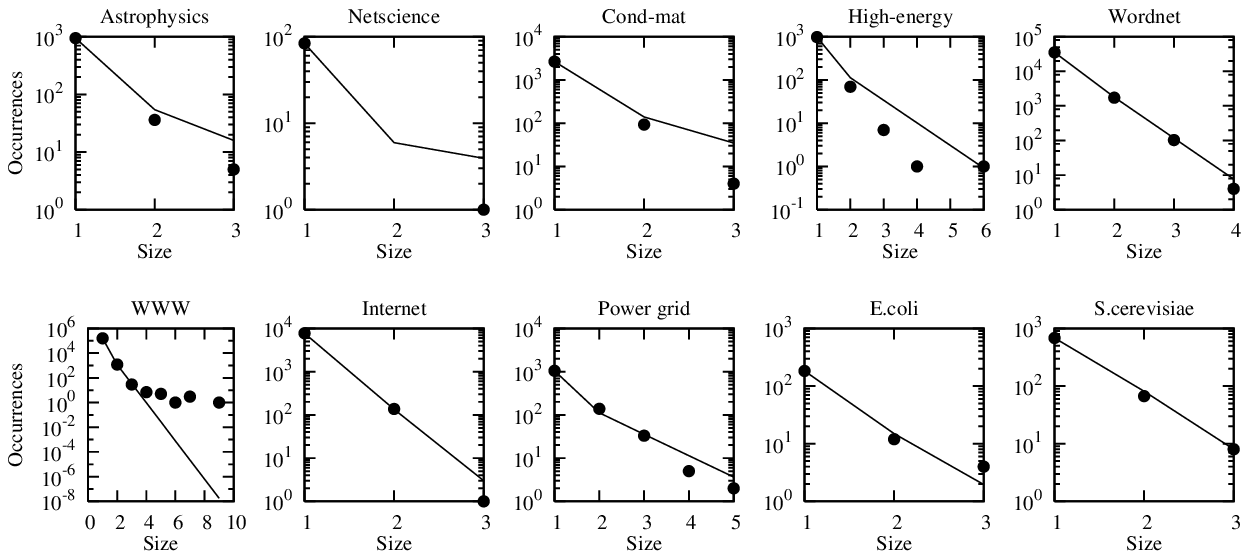}}
  \vspace{0.3cm} \subfigure[\,Number of
  handles.]{\includegraphics[width=0.9\linewidth]{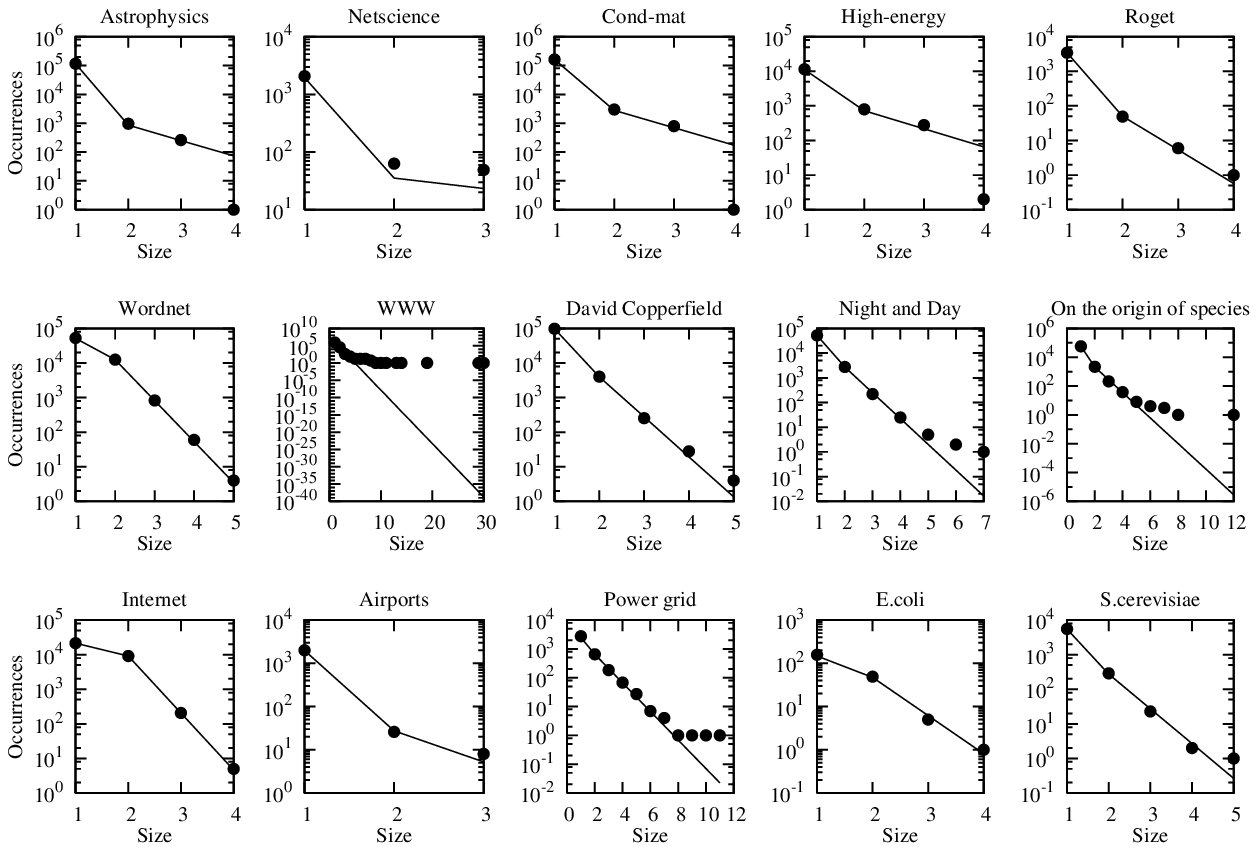}}
  \caption{The distributions shown in (a), (b), and (c) correspond to
  the most significant data (each distribution have at least three
  points). Points correspond to the real data, and the solid lines
  correspond to the theoretical predictions.}  \label{fig:theory}
\end{figure*}

Going back to the analysis presented in Section~\ref{sec:stat}, we
applied those theoretical developments to the considered real-world
networks.  We obtained their degree-degree correlations and computed
the expected number of cords, tails, and handles in function of their
sizes by Equations (\ref{eq:cords}),~(\ref{eq:tails}),
and~(\ref{eq:handles}), respectively. The number of rings was not
taken into account because of their very low probability to appear in
real-world networks. The results concerning the theoretical analysis
are shown in Figure~\ref{fig:theory}. The cases not shown are those
that have all chains smaller than 2. Due to the low probability of
finding cords in networks, only three networks are shown
(Figure~\ref{fig:theory}(a)), namely: cond-mat, high-energy
collaborations and the Wordnet. The theoretical prediction does not
work well for these networks, except for the Wordnet, predicting less
cords than those found in the real networks.  An opposite situation
was found for the number of tails and handles, shown in
Figure~\ref{fig:theory} (b) and (c) respectively. However, there are
more larger tails and handles in the real-world networks than
predicted by theory, except for Astrophysics, cond-mat, and
high-energy collaboration networks.

Despite the fact that, for some cases, the number of small cords,
tails, and handles of the real-world networks were far from the values
obtained from their respective randomized counterparts (see
Figure~\ref{fig:zscore}), the theoretical results were accurate for
several cases, except for astrophysics (handles), netscience (tails),
cond-mat (cords and handles), high-energy (cords, tails, and handles),
WWW (tails and handles), the book On the origin of species (handles),
and power grid (handles) (see (Figure~\ref{fig:theory}).

\subsection{Analysis of incomplete networks} \label{sec:incompleteness}

\begin{figure}
  \includegraphics[width=0.8\linewidth]{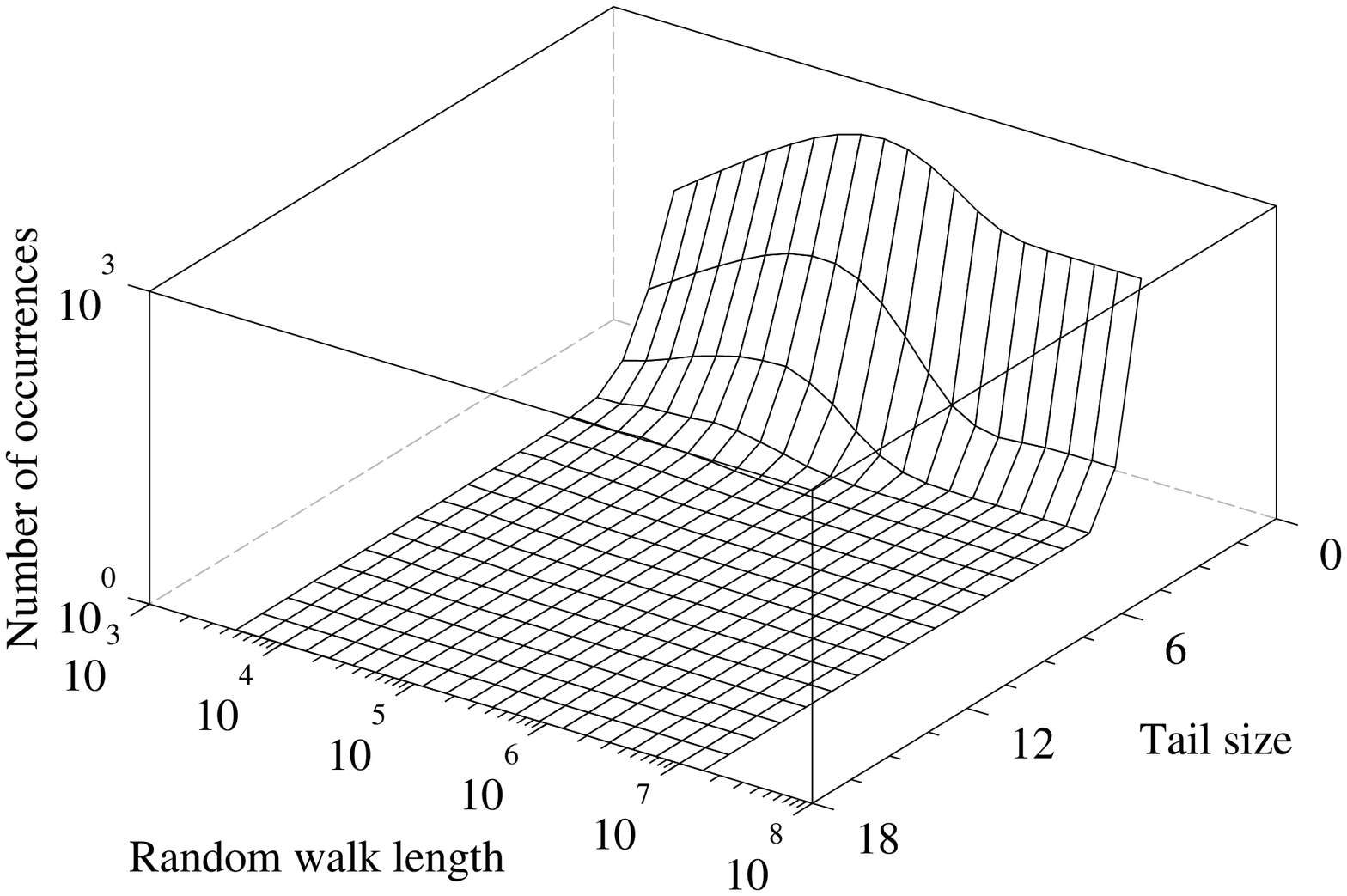} (a) \vspace{0.6cm} \\
  \includegraphics[width=0.8\linewidth]{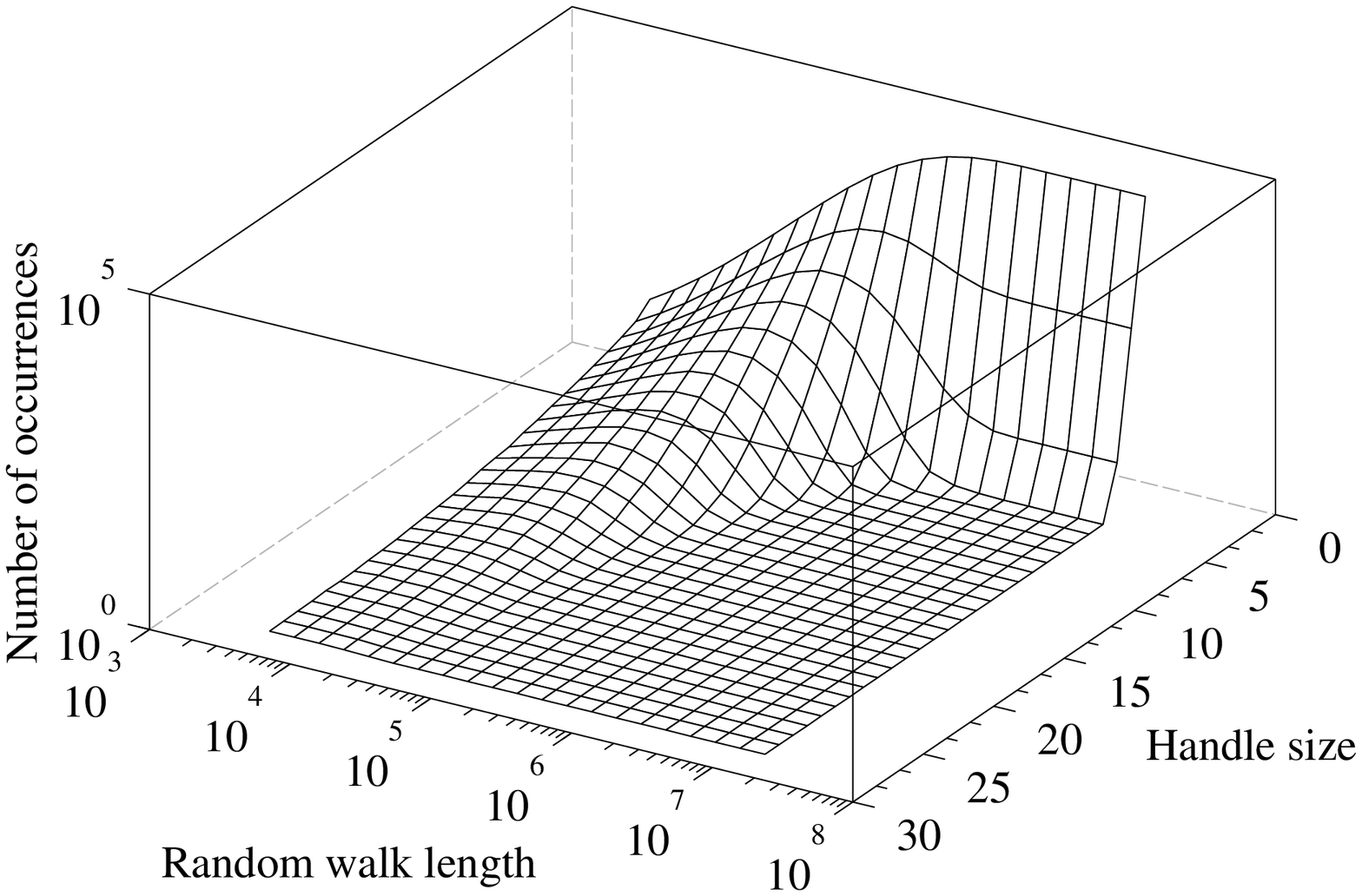} (b) \vspace{0.6cm}\\
  \includegraphics[width=0.8\linewidth]{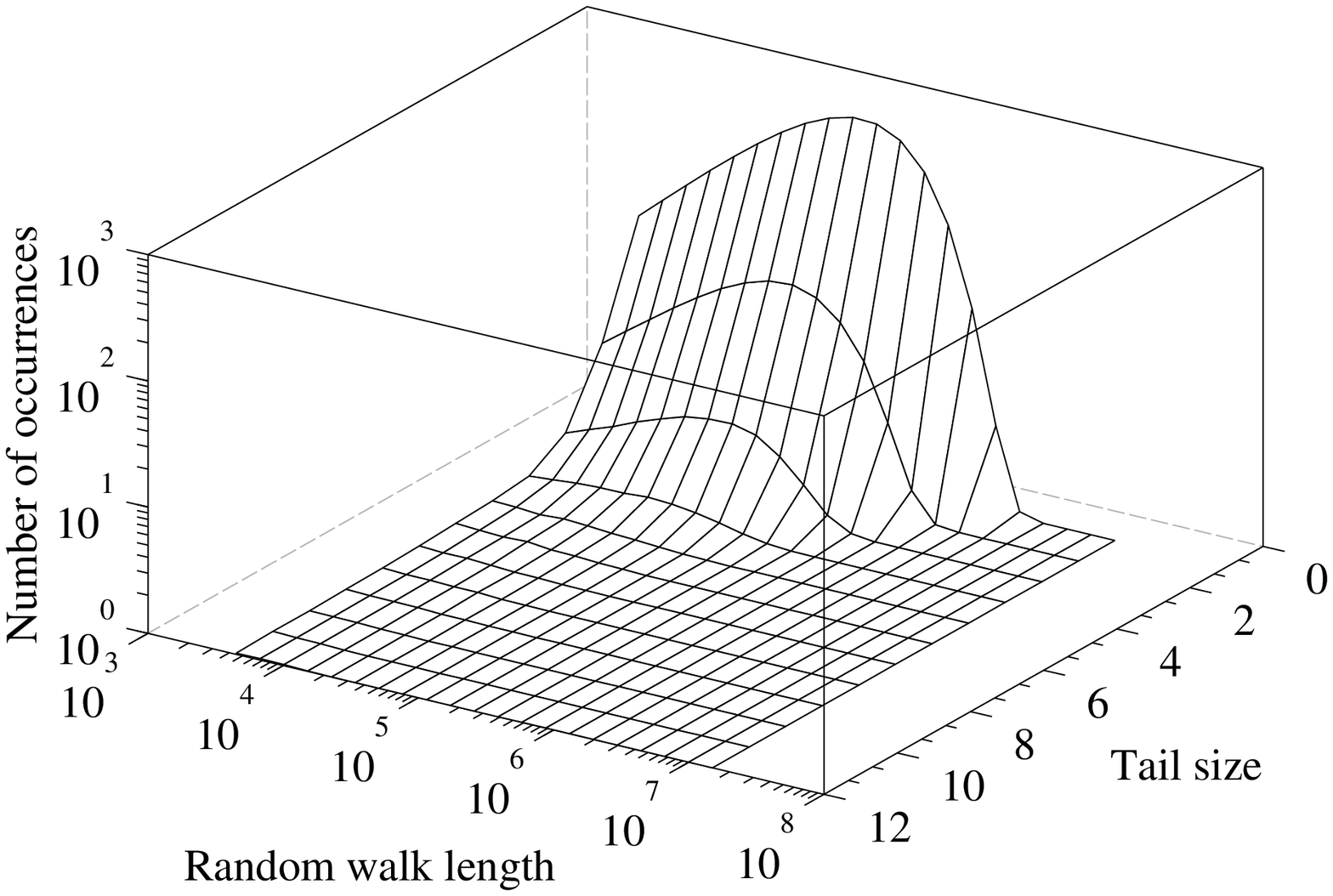} (c) \vspace{0.6cm} \\
  \includegraphics[width=0.8\linewidth]{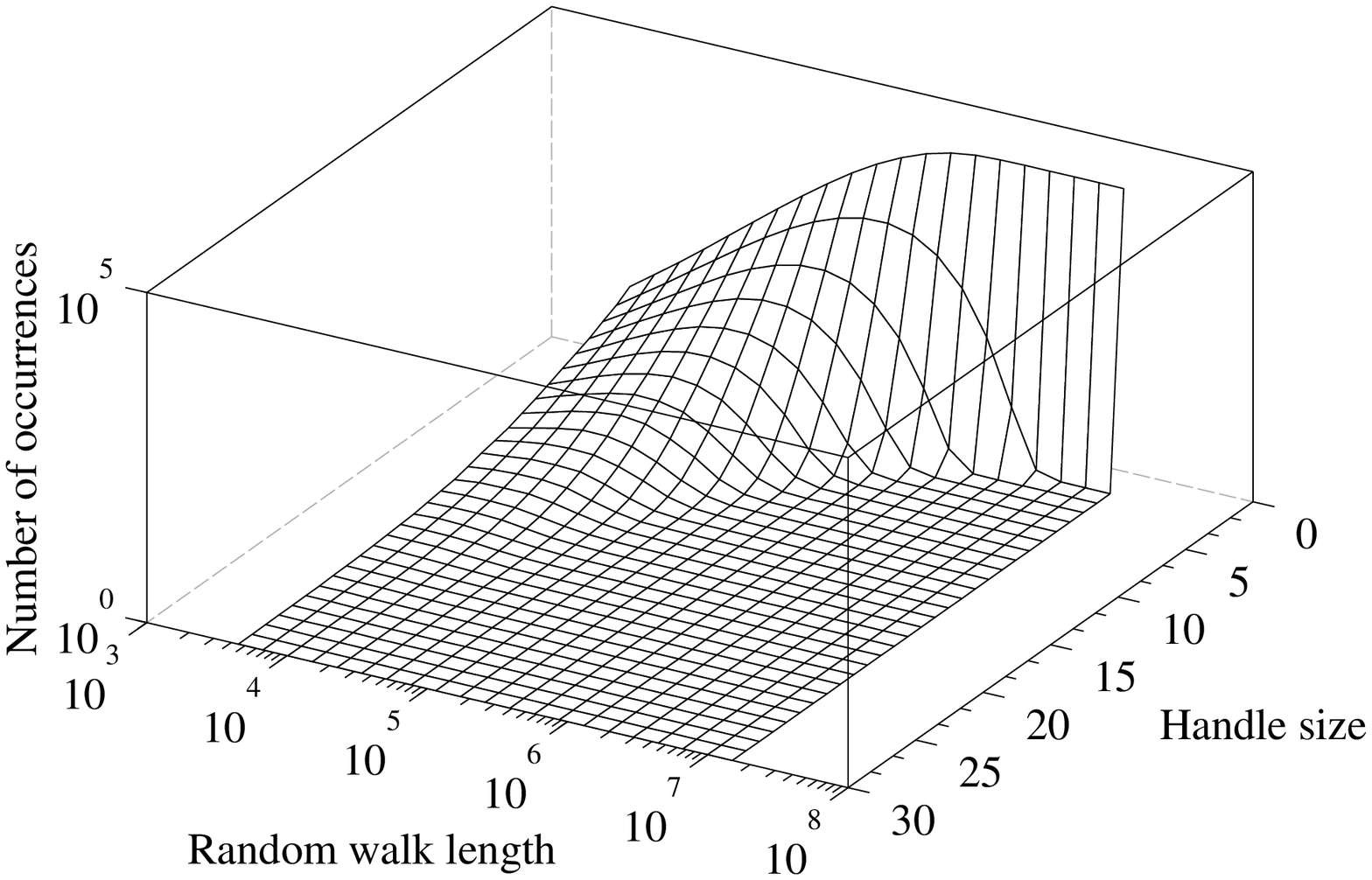} (d)\\
  \caption{Figures (a) and (b) present the number of tails and handles
  of different sizes in the Erd\H{o}s-R\'{e}nyi model,
  respectively. Figures (c) and (d), on the other hand, present the
  number of tails and handles for the Barab\'asi and Albert scale-free
  model, respectively.  Each point in the mesh grid is the average
  considering 1\,000 realizations of each random walk.}
  \label{fig:incompleteness}
\end{figure}

In order to investigate the possibility that incomplete networks
presents many tails and handles, we sampled two theoretical network
models, namely Erd\H{o}s-R\'{e}nyi model (ER)~\cite{Erdos-Renyi:1959}
and Barab\'asi and Albert scale-free model (BA)~\cite{Barabasi:99} by
performing random walks~\cite{noh2004,costa2007ecn}, and analyzing the
corresponding distributions of tails and handles.  The ER and BA
models included 100\,000 vertices with average degree 6. The results
of the random walks in these theoretical networks are shown in
Figure~\ref{fig:incompleteness}. Each point of the mesh grid is the
average value considering 1\,000 realizations.

For the ER and BA models the results are very similar, with the
difference that the tails tend to vanish with larger random walks
(almost $10^7$ steps) in the BA model. This is not the case for the ER
network because its original structure already had vertices with unit
degree. Therefore, this network already had small tails (size 1 and
2). Conversely, BA networks of average vertex degree 6 do not have
tails, and with large random walks these structures tend to vanish.

The results from Figure~\ref{fig:incompleteness} clearly indicates
that there are many large tails and handles for both models when the
random walks are relatively short. As the size of random walks are
increased, the number of large tails and handles tend to decrease, but
the number of small tails and handles increases, because with large
random walks the probability of breaking large tails and handles in
smaller parts is increased. As the length of the random walks increase
further, the large tails and handles tend to vanish, and the original
networks are recovered.

\section{\label{sec:conc}Conclusions}

One of the most important aspects characterizing different types of
complex networks concerns the distribution of specific connecting
patterns, such as the traditionally investigated motifs.  In the
present work we considered specific connecting patterns including
chains of articulations, i.e.\ linear sequences of interconnected
vertices with only two neighbors.  Such a new type of motifs has been
subdivided into cords (i.e.\ chains with free extremities), rings
(i.e.\ chains with no free extremities but disconnected from the
remainder of the network), tails (i.e.\ chains with only one free
extremity) and handles (i.e.\ chains with no free extremity). By
considering a large number of representative theoretical and
real-world networks, we identified that many specific types of such
networks tend to exhibit specific distribution of cords, tails, and
handles. We provide an algorithm to identify such motifs in generic
networks. Also, we developed an analytical framework to predict the
number of chains in random network models, scale-free network models
and real-world networks, which provided accurate approximations for
several of the considered networks. Finally, we investigated the
presence of chains by considering Z-score values (i.e.\ comparing the
presence of chains in real networks and the respective random
counterparts). The specific origin of handles and tails are likely
related to the evolution of each type of network, or incompleteness
arising from sampling. In the first case, the handles and tails in
geographical networks may be a consequence mainly of the chaining
effect obtained by connecting vertices with are spatially
near/adjacent one another. In the second, we showed that incomplete
sampling of networks by random walks can produce specific types of
chains.

All in all, the results obtained in our analysis indicate that handles
and tails are present in several important real-world networks, while
being largely absent in the randomized versions and in the considered
theoretical models. The study of such motifs is particularly important
because they can provide clues about the way in which each type of
network was grown.  Several future investigations are possible,
including the proposal of models for generation of networks with
specific distribution of handles and tails, as well as additional
experiments aimed at studying the evolution of handles and tails in
growing networks such as the WWW and the Internet.

\begin{acknowledgments}

The authors thank Lucas Antiqueira for providing the books
networks. Luciano da F. Costa thanks CNPq (301303/06-1) and FAPESP
(05/00587-5); Francisco A. Rodrigues is grateful to FAPESP
(07/50633-9); Paulino R. Villas Boas is grateful to CNPq
(141390/2004-2); and Gonzalo Travieso is grateful to FAPESP (03/08269-7).

\end{acknowledgments}

\bibliography{chains}

\end{document}